%% file: IEEE_TAP.tex
\documentclass[journal]{IEEEtran}
\usepackage{xcolor}
\usepackage[pdftex]{graphicx}
\graphicspath{{../pdf/}{../jpeg/}}
\DeclareGraphicsExtensions{.pdf,.jpeg,.png}
\usepackage{graphicx}
\usepackage{hyperref}
\hypersetup{
    colorlinks=true,
    linkcolor=blue,
    filecolor=blue,      
    urlcolor=black,
    citecolor =blue,
    pdftitle={Design of Perfect anomalous reflector},
    pdfpagemode=FullScreen
    }
\usepackage[cmex10]{amsmath}
\usepackage{array}
\usepackage{textcomp}
\usepackage{gensymb}
\usepackage{multirow}
\usepackage{multicol}
\usepackage{stfloats}
\usepackage{cite} 
\usepackage[nolist]{acronym}

\begin{document}

\title{Modeling RIS from Electromagnetic Principles to \\Communication Systems--Part I: Synthesis and Characterization of a Scalable Anomalous Reflector}

\author{
Sravan~K.~R.~Vuyyuru,~\IEEEmembership{Member,~IEEE,}
Le~Hao,~\IEEEmembership{Student Member,~IEEE,}
Markus~Rupp,~\IEEEmembership{Fellow,~IEEE,}\\
Sergei~A.~Tretyakov,~\IEEEmembership{Fellow,~IEEE,}
and Risto~Valkonen,~\IEEEmembership{Member,~IEEE}

\thanks{This work was supported by the European Union’s Horizon 2020 MSCA-ITN-METAWIRELESS project, under the Marie Skłodowska-Curie grant agreement No 956256. \textit{(Corresponding author: Sravan~K.~R.~Vuyyuru)}} 
\thanks{S.~K.~R. Vuyyuru is with Nokia Bell Labs, Karakaari 7, 02610 Espoo, Finland and the Department of Electronics and Nanoengineering, School of Electrical Engineering, Aalto University, 02150 Espoo, Finland (e-mail: sravan.vuyyuru@nokia.com; sravan.vuyyuru@aalto.fi).}
\thanks{L. Hao, and M. Rupp are with TU Wien, Gusshausstrasse 25, 1040 Vienna, Austria. (e-mail: {le.hao, markus.rupp}@tuwien.ac.at).}
\thanks{R. Valkonen is with Nokia Bell Labs, Karakaari 7, 02610 Espoo, Finland (e-mail: risto.valkonen@nokia-bell-labs.com).}
\thanks{S.~A. Tretyakov is with the Department of Electronics and Nanoengineering, School of Electrical Engineering, Aalto University, 02150 Espoo, Finland (e-mail: sergei.tretyakov@aalto.fi).}
}

\maketitle\begin{abstract}
This work aims to build connections between the electromagnetic and communication aspects of Reconfigurable Intelligent Surfaces (RIS) by proposing a methodology to combine outputs from electromagnetic RIS design into an RIS-tailored system-level simulator and a ray tracer. In this first part of the contribution, a periodic anomalous reflector is designed using an algebraic array antenna scattering synthesis technique that enables electromagnetically accurate modeling of scattering surfaces with both static and reconfigurable scattering characteristics. The multi-mode periodic structure, capable of scattering into several anomalous angles through manipulation of reactive loads, is then cropped into finite-sized arrays, and the quantization effects of the load reactances on the array scattering are analyzed. An experimental anomalous reflector is demonstrated with a comparison between simulated and measured scattering performance. In the second part, the simulated receiving and transmitting scattering patterns of the anomalous reflector are utilized to build an electromagnetically consistent path loss model of an RIS into a system-level simulator. Large-scale fading is analyzed in simple scenarios of RIS-assisted wireless networks to verify the communication model, and an indoor scenario measurement using the manufactured anomalous reflector sample to support the simulation analysis. After verifying the connections between electromagnetic and communication aspects through simulations and measurements, the proposed communication model can be used for a broad range of RIS designs to perform large-scale system-level and ray-tracing simulations in realistic scenarios.
\end{abstract}


\IEEEpeerreviewmaketitle

%=== SECTION I: Introduction =================================================

\input{Acronyms.tex}
\section{Introduction}\label{sec:Intro}

\IEEEPARstart{R}{econfigurable} Intelligent Surface (RIS) is an emerging technology with diverse applications in next-generation wireless communication networks, owing to its capacity for dynamic control and optimization of the propagation environment~\cite{Smart_Radio_Environments,MarcoCommModelsRIS}. The rapid evolution driven by escalating performance demands of coverage in blocked line-of-sight scenarios has prompted the exploration of RIS. Specifically, RIS owns a primary functionality of deflecting incoming signals towards arbitrary non-specular directions by violating the conventional reflection law, anointed as anomalous reflectors. Modeling RIS in terms of the optimized distribution of the local reflection coefficient defined for each array element~\cite{Najafi21,Ozdogan20} and integrating these extensive codebook discrete sets into network simulators raises complexity. Besides, that conventional approach is based on inaccurate physical models of non-uniform reflectors, often resulting in practically not realizable requirements on the response of the  individual elements. The RIS develops a collective response defined by many strongly interacting controllable cells. 

\begin{figure*}[t!]
\begin{center}
\includegraphics[width=7in,height=1.5in]{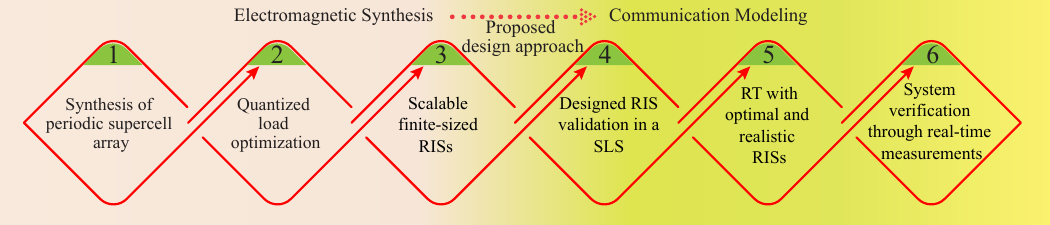}
\caption{Flowchart depicting the proposed RIS modeling process from \ac{EM} synthesis to communication modeling. SLS: System-level simulator; RT: Ray tracing}\label{fig:RIS_Theme}
\end{center}
\end{figure*}

For example, the authors of~\cite{Tang2021,Tang2022} proposed a system model for RIS-assisted wireless communications from the \ac{EM} perspective, based on the local reflection coefficient model and the notion of gains of the individual array elements. In the assumption that the reflection phase distribution is optimal for reflection into the desired direction, simple path-loss formulas have been obtained, but more studies are needed to find out how to incorporate these models into ray-tracing system-level simulators. Moreover, practical realizations of these local reflection coefficients are challenging (if at all possible). In \cite{Sergei2023}, simple link-budget formulas were derived from electromagnetic solutions for optimally functioning anomalous reflectors, where the collective nature of reflection phenomena was accounted for, and the notion of the local reflection coefficient was not used. However, it is still necessary to study if and how such optimal and scanning anomalous reflectors can be practically realized and, on the other hand, how to properly model these RISs in system-level simulators. 

In filling these research gaps, this study aims to build a bridge between \ac{EM} principles and design methods and system-level models of communication systems that incorporate RIS, as illustrated in Fig.~\ref{fig:RIS_Theme}. Our study integrates two principal directions: one delves with the \ac{EM} perspective, while the other focuses on the communications part. In  the first part, we design, characterize, and experimentally test an anomalous reflector based on a periodic 'supercell' structure that can be set to reflect the incident waves into a relatively small number of discrete directions. First, we design a periodic linear array of generic unit cells. Subsequently, we scale this supercell to medium-size square finite arrays using the array factor approximation to study and evaluate the scalability of the anomalous reflector design, eliminating the need for separate designs for each finite array size. We validate these designs through numerical simulations to demonstrate their stability against physical scaling issues. To further address practical reflector implementations, we investigate the impacts of spatial discretization of the anomalous reflectors against ideal continuous reflector, and load impedance quantization against continuous load tuning. Finally, we manufacture a static anomalous reflector prototype to assess the implications of cropping the periodic design of an anomalous reflector into a practical-size array.

Consequently, in the second part, we discuss methodologies to pass the macroscopic parameters of the designed RIS to network simulators to facilitate a rapid design process in developing an appropriate level of abstraction concerning the macroscopic parameters to account for both an accurate EM behavior while maintaining low system-level modeling complexity for efficient network planning through reliable propagation channel modeling with reduced computational complexity. It benefits in developing robust RIS-assisted wireless networks, avoiding local reflection coefficients of multiple cells, and conducting extensive computational characterization and measurements of \ac{EM} patterns of manufactured RIS. As a result, to realize a RIS, we replace heavy computations required for channel optimization with simple algebraic optimization in modeling a collective response of a periodic supercell structure and enhances rapid input data loading for network simulators where RIS is modeled by its geometrical area and the angles of arrival and reflection. Part~II concentrates mainly on the communication-related aspects and analyzes the relation between the \ac{EM} and communication models by incorporating realistic and accurate scalable RIS \ac{EM} simulation parameters into network simulators to verify the large-scale fading of RIS-assisted links. We estimate the overall system performance with the proposed technique as a final step in Fig.~\ref{fig:RIS_Theme} and assist in deploying electrically large RISs into complex environments without ample measurements. 
 
The remainder of this paper is organized as follows. In Section~\ref{sec:RISSynthesis}, we give a motivation to Part~I of this study and outline the design synthesis procedure for  periodically loaded anomalous reflectors. Section~\ref{sec:RISfloquet} explains the theoretical problem formulation for a large supercell array allowing multiple Floquet harmonics. Section~\ref{sec:RISdesign} introduces the developed multi-mode optimized anomalous reflector using the scattering synthesis theory. Section~\ref{sec:RISdesign_cont} and Section~\ref{sec:RISdesign_quan} present the implications of discretization, regarding both spatial discretization of the surface and load quantization. In Section~\ref{sec:RISExperimental}, we demonstrate experimental results for a manufactured passive fixed anomalous reflector. Finally, conclusions for this part are drawn in Section~\ref{sec:conclusion}.

%=== SECTION II: RIS design =================================================

\section{Scattering Synthesis for Multi-mode Anomalous Reflector}\label{sec:RISSynthesis}

\begin{figure*}[t]
\begin{center}
\includegraphics[width=7in,height=2.2in]{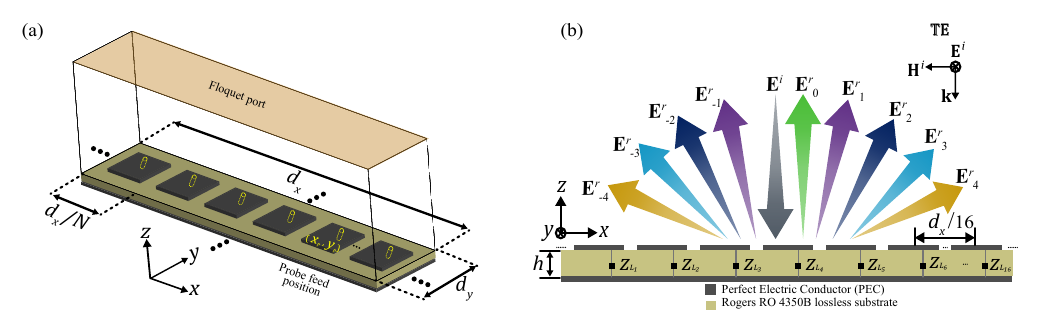}
\caption{(a) Schematic representation of the supercell of reflector metasurface design geometry with $N$ elements (b) Side view of the supercell. The 2-D Schematic sketch of the multimode anomalous reflector with nine propagating harmonics for normal incidence in TE polarization.}
\label{FigS:Schematic_multiref}
\end{center}
\end{figure*}

Advanced methods of designing high-efficiency anomalous reflectors using periodic metasurfaces~\cite{diaz2017generalized,wong_prx2018,MacroscopicARM2021,kwon_ieeejawpl2018,asadchy_prx2017,asadchy2016perfect,vuyyuru,movahediqomi2023comparison,vuyyuru2023finite,li2023tunable} and 
metagratings~\cite{MG_YounesAlu,popov2019Feb,design_MG_LPA_popov,MultipleChannel2020Wang,Dual_Pol_MG_Epstein} have been developed. However, the research in designing efficient and scanning RIS with anomalous reflection is still evolving. The prior research focused on designing an anomalous reflector based on the generalized law of reflection~\cite{yu_science2011} or the phased-array principle~\cite{huang2008}. These reflectors exhibit a local reflection phase gradient along the surface. Nevertheless, it is known that the scattering power efficiency of these conventional phase-gradient reflectors gradually declines as deflection angles increase, e.g.~\cite{diaz2017generalized}. 

This difficulty has been addressed by employing a periodic reflector configuration, which allows for manipulating the propagating Floquet modes by optimizing evanescent fields near the reflector~\cite{asadchy_prx2017}. In~\cite{wong_prx2018}, a periodic reflector is devised using a supercell period comprising two meta-atoms, allowing for independent control over two propagating Floquet modes. In~\cite{asadchy_prx2017}, individual Floquet modes are conceptualized as channels in an $N$-port device by controlling the structural periodicity and defining multichannel metasurfaces. However, that design requires multiple component dimensions to achieve an optimized scattering matrix, making it impractical to deploy in a real-time environment in realizing multiple and reconfigurable reflection angles within a single design. Additionally, optimizing the meta-atoms of the supercell through full-wave \ac{EM} simulations proves inefficient, demanding substantial computational resources and prolonged optimization times. An arithmetic optimization method has been developed using array antenna scattering synthesis for periodic reflectors comprising reactively loaded patch elements in a supercell~\cite{vuyyuru}. The method transfers the load optimization to the circuit-based algebraic optimization domain, offering a numerically efficient solution compared to computation-intensive full-wave \ac{EM} simulations. Yet, this paper confines the reflector design to generating beams into discrete angles aligned with  single supported Floquet harmonic, restraining its scalability for practical deflectors suitable for use as RIS. 

Here, we propose a scattering synthesis methodology in designing a multi-mode periodic anomalous reflector with a multi-wavelength supercell dimension, demonstrating numerically efficient optimization to maximize the deflection efficiencies for all supported multiple perfect anomalously scattering Floquet modes. The background idea is to then extend the periodic models to finite-sized arrays to study the scalability of the multi-mode reflector design method.
Optimizing periodic arrays rather than finite arrays benefits from avoiding global optimization and enabling rapid algebraic design for a single supercell. Finally, we conduct measurements utilizing a fixed anomalous reflector prototype, providing a comparison between the simulated and experimental results.

Our particular design goal is to synthesize a planar, periodic reflector supporting scattering into multiple anomalous directions and simultaneously suppressing specular reflections when illuminated at a fixed incident angle. We achieve this goal by applying the Floquet theory to an electrically large supercell of an infinite periodic array and optimizing the reactive loads of the elementrs in this supercell for each desired scattering mode. First, we conceptually study and characterize the periodic anomalous reflectors made from generic patch antenna unit cells, using Floquet harmonics to control the allowed desired reflection lobes. Then, this electrically large supercell with periodic (infinite) structure is cropped into a finite antenna array to study the characteristics of anomalous reflectors used as RIS in complex \ac{EM} environments.

\subsection{Scattered fields of Floquet Harmonics}\label{sec:RISfloquet}

Consider a rectangular supercell of dimensions $d_x$ and $d_y$, with periodic boundaries in $x$- and $y$-directions, and its radiating surface lying on the $xy$-plane of a 3-dimensional coordinate system. This supercell of smaller unit cells consists of $N$ identical patches on a planar substrate, backed with a perfect electric conductor (PEC) ground plane. Each patch is loaded with a variable reactive impedance load ${Z}_L$, allowing configuring reflection into several directions by reconfiguration of the set of loads. The infinite array of such supercells is illuminated by a plane wave with a constant $E$-field magnitude and transverse electric (TE) polarization
\begin{equation}
\mathbf{E}^i=\mathbf{E}^i_0
e^{+j\mathbf{k}^i\cdot\mathbf{r}},
\label{eq1}
\end{equation}
where $\mathbf{r}=\hat{x}x+\hat{y}y+\hat{z}z$, and the incident wave vector, $\mathbf{k}^i$, 
points in the incoming wave direction $(\theta^i,\phi^i)$. The total scattered $E$-field amplitude, $\mathbf{E}^s_m(\mathbf{Z}_L)$, scattered by a periodic anomalous reflector into the $m$th propagating Floquet mode in $(\theta^r_m,\phi^r_m)$ is given by the sum of zero-current scattering and port-current scattering~\cite{vuyyuru}:
\begin{equation}
\mathbf{E}^s_m(\mathbf{Z}_L)=
\mathbf{E}^s_m(\mathbf{Z}_L=\infty)
-\sum_{n=1}^N\frac{k\eta I_{Ln}(\mathbf{Z}_L)}{2d_xd_yk_{zm}}
\mathbf{h}^\text{el}_n(\theta^r_m,\phi^r_m),
\label{eq2}
\end{equation}
where $k$ is the free-space wavenumber, $\eta\approx 377~\Omega$ is the free-space intrinsic impedance, and $k_{zm}=k\cos\theta^r_m$. $\mathbf{E}^s_m(\mathbf{Z}_L=\infty)$ is the scattered field from the periodic array for the $m$th propagating Floquet mode when the load terminals are open-circuited. This data can be acquired directly from open-circuited RX-mode simulation as an input for algebraic optimization. $I_{Ln}(\mathbf{Z}_L)$ and $\mathbf{h}^\text{el}_n(\theta^r_m,\phi^r_m)$ are the load port current dependent on load impedance ${Z}_L$ and the element vector effective height for predetermined $(\theta^r_m,\phi^r_m)$, respectively, with $N\times 1$ column vector at the $n$th port~\cite{vuyyuru,EffectiveheightDo-Hoon}. The element vector effective height $\mathbf{h}^\text{el}_n$ data is acquired from transmitting (TX) simulation. The load port currents obtained from the preliminary data from simulations, based on~\cite{vuyyuru}.

We aim to optimize an anomalous reflector to support multiple Floquet modes with scanning only in the plane of illumination (here, the $xz$-plane). Therefore, our supercell comprises only a single row of loaded patches, as depicted in Fig.~\ref{FigS:Schematic_multiref}. The $x$-component of the $m$th Floquet harmonic for periodical arrays is defined as the tangential wavenumber of the reflected harmonic, $k_{xm}$, along the $x$-axis as
\begin{equation}
k_{xm}=k_{x0}+\frac{2m\pi}{d_x},
\label{eq3}
\end{equation}
where $k_{xm}=k\sin\theta^r_m$ and $k_{x0}=k\sin\theta^i$. The supercell size $d_x$ determines the ability to scatter into multiple reflection angles, depending on $\theta^i$ and  $\theta^r$. The reflection angle for the $m$th Floquet harmonic $\theta^r_m$ is determined by
\begin{equation}
\sin\theta^r_m=\sin\theta^i+\frac{m\lambda}{d_x},
\label{eq4}
\end{equation}
which needs to be real-valued in order to support the desired Floquet harmonics to scatter, thereby limiting the number of supported re-radiation modes. The additional higher-order (evanescent) modes do not significantly contribute to the far field as their propagation constants in the normal direction are imaginary. Controlling the propagating Floquet modes via circuit-based algebraic optimization through the controllable reactive loads enables accurate prediction of reflection amplitudes, circumventing the need for cumbersome full-wave \ac{EM} simulations.

\subsection{Spatial Discretization of Reflector Surface}\label{sec:RISdesign}
 
\begin{figure}[tbh!]
\begin{center}
\includegraphics[width=3.45in,height=2.7in]{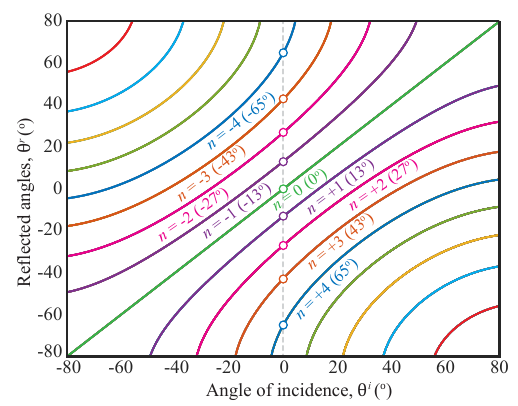}
\caption{The relation between the incidence angle $\theta^i$ and deflection angle $\theta^r$ varies across different Floquet modes, given by $d_x=4\lambda/\sin\theta^r$ periodicity.}\label{fig:Modes}
\end{center}
\end{figure}

A TE-polarized anomalous reflector for  multiple reconfigurable reflected Floquet modes is designed under normal incidence $(\theta^i\!=\!0^\degree)$ at the design frequency $26$~GHz. For the periodic arrays, we choose the supercell size based on the desired reflection angles. In this work, we allow propagating Floquet harmonics up to fourth order. From the basics of the Floquet-Bloch theory, we choose the supercell dimensions to be $d_x\!=\!4\lambda/\sin\theta^{r,max}$. We set the maximum desired reflection angle $\theta^{r,max}\!=\!65^\degree$, yielding $d_x\!=\!4.4135\lambda$ ($50.89$~mm). Satisfying $|k_{xm}|<k$, there are nine propagating harmonics with $m\!=\!0$, $\pm 1$, $\pm 2$, $\pm 3$, $\pm 4$, as depicted in Fig.~\ref{FigS:Schematic_multiref}, and all the other modes are evanescent. Figure~\ref{fig:Modes} shows the relation between the incidence angle $\theta^i$ and deflection angle $\theta^r$ across the available Floquet modes. These nine propagating Floquet modes have nominal rescattering directions at $\theta^r\!=\!0\degree$, $\pm 13\degree$, $\pm 27\degree$, $\pm 43\degree$, and $\pm 65\degree$ deflection angles from the normal.

We still need to determine the required number of patch elements, $N$, within the supercell. Based on \cite{popov2019Feb}, the multimode anomalous reflector structure needs two passive scatterers to control one Floquet mode coefficient, meaning that we would need eighteen loaded scatterers to fully control nine Floquet modes. Meanwhile, the optimization of scattering synthesis in~\cite{vuyyuru} demonstrated that nearly perfect anomalous reflection could be achieved with less than two scatterers per half wavelength. We choose $N\!=\!16$ to have a convenient power-of-two subdivision of our supercell. We thus consider a linear subarray comprising sixteen identical patches with $d_x/N \!= d_y\! =\! 0.2758\lambda$ ($3.1806$~mm at 26~GHz) unit cell spacing. 

\begin{table}[thb!]
\renewcommand{\arraystretch}{1.6}
    \begin{center}
    \caption{Dimensions of the supercell array}
    \label{tab:dimentions}
    \footnotesize
    \begin{tabular}{|c|c|} \hline
    Parameters & Supercell dimensions \\ \hline
    \multirow{2}{*}{Overall dimension $(d_x\times d_y)$} & $4.4135\lambda\times0.2758\lambda$ \\ 
    & ($50.89$~mm $\times~3.1806$~mm) \\ \hline
    \multirow{2}{*}{Square patch dimension} & $0.2328\lambda$ \\ 
    & ($2.6838$~mm) \\ \hline
    Substrate height ($h$) & $0.338$ mm \\ \hline
    Probe position $(x_p,y_p)$ & $(0,1.068)$ mm \\ \hline
    Total loaded elements $(N)$ & $16$ \\ \hline
    \end{tabular}
    \end{center}
\end{table}

We begin our investigation by designing a unit-cell resonant patch using the periodic dimensions $d_x/N$ by $d_y$, supporting only a single Floquet mode alone. This step is a design choice made to implement a simple internal subdivision geometry of the 16-element supercell. The single-patch unit cell simulations are made using CST Microwave Studio simulator. We adopt the approach outlined in~\cite{vuyyuru} for determining the dimensions of the patch unit cell. We use a linearly polarized PEC square patch antenna at 26~GHz design frequency with  $50~\Omega$ matching impedance. In simulations, the substrate material is lossless RO4350B with a relative permittivity $\epsilon_r\!=\!3.66$. The patch dimensions along with the feeding position optimized for $50~\Omega$ load impedance for the desired frequency are presented in Table~\ref{tab:dimentions}. This single-patch design is replicated into a linear periodic supercell of sixteen patches, allowing individual reactive load terminations replacing the antenna feeds. In preparation for the supercell optimization, we define the fitness function that maximizes the power reflection efficiency $\zeta$ for each desired propagating Floquet mode~\cite{vuyyuru}. The loads are optimized following Section~\ref{sec:RISfloquet} to find the final sets of reactive loads to deflect normally incident waves into each of the supported anomalous reflection modes with indices $m$. 

\subsection{Continuous Phase Periodic Supercell Optimization}\label{sec:RISdesign_cont}

First, we optimize these periodic arrays to fully reflect incident waves from $\theta^i\!=\!0$ to angles $\theta^r\! =\! 13\degree$, $27\degree$, $43\degree$, and $65\degree$ without setting any limitations on the individual load reactances. The optimized (rounded) load reactance values and the respective achieved scattering mode efficiency values $\zeta$ are reported in Table~\ref{tab:optquantizedload}. With unrestricted load variables optimization, the supercell with sixteen load-adjusted patches can accomplish nominally perfect efficiencies for all the propagating Floquet harmonics with only algebraic optimization. 

\begin{figure}[tbh!]
\begin{center}
\includegraphics[width=3.48in,height=2.22in]{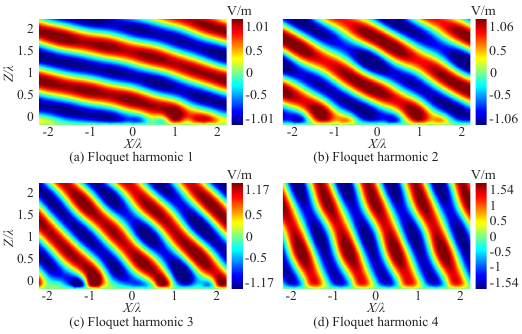}
\caption{The $y$-component of the scattered $E$-field \text{Re}\{$E^s_y(x,z)$\} produced by the TE polarized multimode reflector controlling propagating modes from $1$ to $4$ for 13\textdegree, 27\textdegree, 43\textdegree and 65\textdegree  deflector, respectively, with a sixteen-patch supercell under a unit-amplitude plane wave illumination at normal incidence.}\label{fig:e_field_scattering}
\end{center}
\end{figure}

\begin{figure*}[tbh!]
\begin{center}	
\includegraphics[width=7in,height=5.5in]{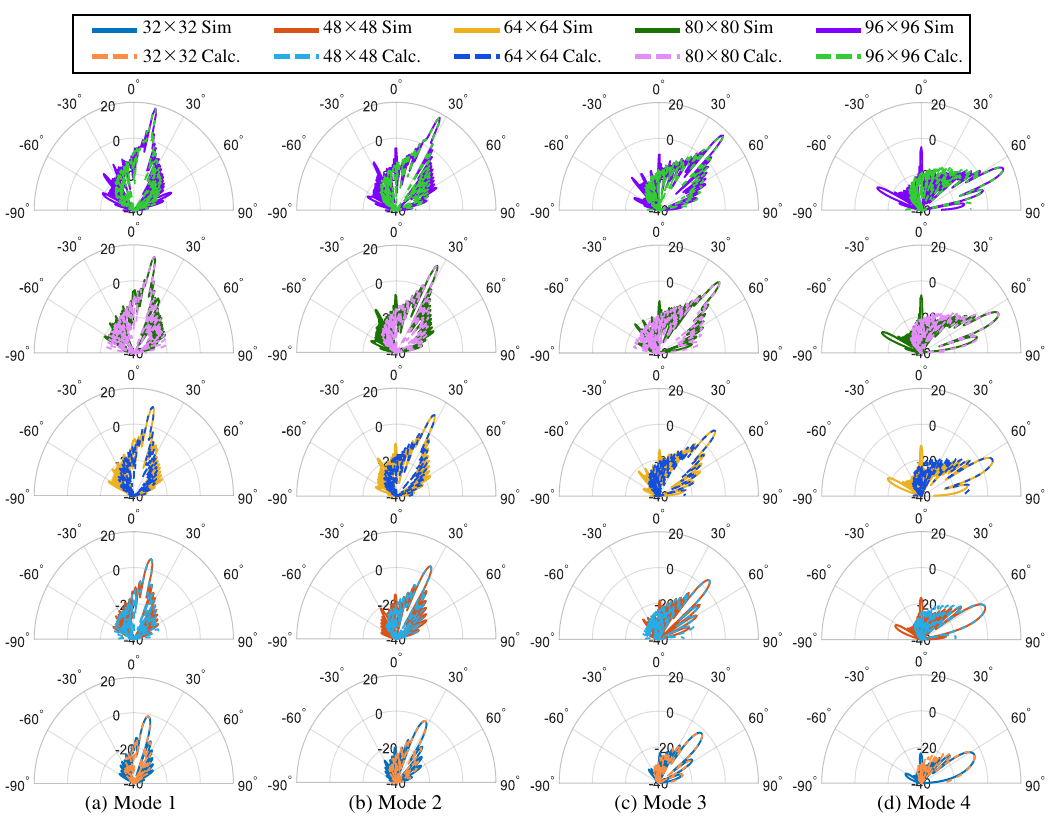}
\caption{Computed far-zone $E$-field employing continuous load variable optimization comparing CST simulated (solid lines) and analytical (dashed lines) fields for five different finite array sizes of $32\times32$, $48\times48$, $64\times64$, $80\times80$, and $96\times96$ for (a) $m=1$, (b) $m=2$, (c) $m=3$, and (d) $m=4$ propagating harmonics, reported in dB(V/m) scale.
}
\label{fig:farfield_cont}
\end{center}
\end{figure*}

Figure~\ref{fig:e_field_scattering} depicts the $y$-component of the reflected E-field snapshot, $E^s_y$, for the optimized modes for $m\!=\!1$ to $m\!=\!4$ with $13\degree$, $27\degree$, $43\degree$, and $65\degree$ reflectors, respectively, illuminated by a plane wave $(\mathbf{E}^i_0\!=\!\hat{y}~\text{V/m})$ at normal incidence. A nearly perfect plane wave propagating in the desired direction can be observed in each case. We observe an enhanced amplitude of the scattered electric field with an increasing deflection angle, as seen in Fig.~\ref{fig:e_field_scattering}, as required for perfect anomalous reflectors. Next we extend our discussion to scalable arrays of manufacturable sizes. Far-zone $E$-fields of finite-size arrays are plotted in Fig.~\ref{fig:farfield_cont} using the array factor approximation by applying an array factor calculation on the ``element pattern'' of the simulated supercell. While the simulation is performed for an infinite array of supercells under periodic boundary conditions, we approximate the resulting radiation characteristics to synthesize a finite-length array of five distinct sizes for practical applications. The five different approximated finite arrays are $32\times32$, $48\times48$, $64\times64$, $80\times80$, and $96\times96$ comprising $1024$, $2304$, $4096$, $6400$, and $9216$ individual scatterers, respectively. The results show a dominant scattering beam into the desired direction while eliminating the specular lobes. 

\begin{table*}[tbh!]
\renewcommand{\arraystretch}{1.2}
\caption{Summary of the optimized load reactances and comparison with the quantized loads.}
\label{tab:optquantizedload}
\centering
\begin{tabular}{| c | c | c | c |} \hline
Floquet mode ($\theta^r$) & Resolution &  Optimized load reactances  $(\Omega)$ & Efficiency, $\zeta$ $(\%)$ \\   \hline
\multirow{5}{4em}{Mode $1$ $(13^{\degree})$} & Continuous & $34, 20, 22, 75, 57, 88, 298, 162, -288, -153, -114, 18, -64, -2, 3, -17$ & 99.6  \\ \cline{2-4}
& $4$ bit  & $-5, 113, -5, 215, 29, 1791, 68, -300, -83, -83, -57, -15, -27, -5, 5, 29$ & 99.4  \\ \cline{2-4}
& $3$ bit  & $-134, 45, -57, -27, 16, -5, -5, 45, 45, -27, 1791, -57, 1791, 113, -134, -134$ & 97  \\ \cline{2-4}
& $2$ bit  & $1791, -57, 1791, 1791, -57, -57, -57, -57, 45, -57, 45, -5, -5, 45, 45, -5$ & 93.1  \\ \cline{2-4}
& $1$ bit  & $45, -57, -57, -57, -57, -57, -57, 45, -57, 45, 45, 45, 45, 45, 45, 45$ & 51.1  \\ \hline
\multirow{5}{4em}{Mode $2$ $(27^{\degree})$} & Continuous & $-1, -364, -66, -18, -44, 108, -28, 323, -2, -359, -77, -12, -44, 107, -27, 319$ & 99.3  \\ \cline{2-4}
& $4$ bit  & $29, 5, 215, 16, -300, 5, -57, -5, 45, -5, 215, 16, -300, -15, -57, 5$ & 99  \\ \cline{2-4}
& $3$ bit  & $113, -5, 1791, -27, -134, -27, 16, -27, 113, -5, 1791, -57, -134, -27, 16, -27$ & 98.1  \\ \cline{2-4}
& $2$ bit  & $-57, -5, -5, 45, 45, 45, 1791, -57, -57, -5, -5, 45, 45, 45, 1791, -57$ & 92.3  \\ \cline{2-4}
& $1$ bit  & $45, -57, -57, -57, 45, -57, 45, 45, 45, -57, -57, -57, -57, 45, 45, 45$ & 47.1  \\ \hline
\multirow{5}{4em}{Mode $3$ $(43^{\degree})$} & Continuous & $-57, 365, -15, -215, 263, -51, 85, 143, -905, -111, 140, -43, 712, -65, -134, 221$ & 99.5  \\ \cline{2-4}
& $4$ bit  & $68, -300, 16, 5, 16, 215, 215, -134, 29, 29, 45, 1791, -83, -5, -15, 113$ & 95.6  \\ \cline{2-4}
& $3$ bit  & $45, -5, 1791, -134, -57, 16, 45, 113, 1791, -57, -27, 45, 45, 1791, -134, -27$ & 95.2  \\ \cline{2-4}
& $2$ bit  & $-57, -5, 45, 45, 1791, -57, -57, 45, -5, 1791, -57, -57, -5, 45, 45, 1791$ & 91.6  \\ \cline{2-4}
& $1$ bit  & $-57, 45, 45, -57, -57, -57, 45, 45, 45, -57, -57, 45, 45, 45, -57, -57$ & 41.2  \\ \hline
\multirow{5}{4em}{Mode $4$ $(65^{\degree})$} & Continuous & $255, -210, 31, 22, 255, -212, 31, 22, 252, -212, 32, 23, 253, -212, 33, 22$ & 99.1  \\ \cline{2-4}
& $4$ bit  & $-300, 16, 29, 215, -300, 16, 29, 215, -300, 29, 29, 215, -300, 29, 29, 215$ & 98.4  \\ \cline{2-4}
& $3$ bit  & $-134, 45, 45, 1791, -134, 45, 45, 1791, -134, 45, 45, 1791, -134, 45, 45, 1791$ & 87  \\ \cline{2-4}
& $2$ bit  & $1791, -57, 45, 45, 1791, -57, 45, 45, 1791, -57, 45, 45, 1791, -57, 45, 45$ & 83.7  \\ \cline{2-4}
& $1$ bit  & $45, -57, -57, 45, 45, -57, -57, 45, 45, -57, -57, 45, 45, -57, -57, 45$ & 36  \\ \hline
\end{tabular}
\end{table*}

In comparison to the numerical EM solver approach using spatially discretized periodic structure, we calculate an analytical reference illustrating ideal anomalous reflectors with 100\% efficiency and continuous surface current distribution. Following~\cite{MacroscopicARM2021}, the analytical approach under the physical optics approximation is employed, suitable for reflectors significantly larger than the wavelength. We consider a planar array with dimensions $a$ and $b$ in the $x$- and $y$-directions, placed on the $xy$-plane and excited by a transverse electric (TE) incident plane wave in the $xz$-plane with the $E$-field ($\mathbf{E}^i = \hat{y}E^i_0 e^{+jk(x\sin\theta^i+z\cos\theta^i)}$). The scattered $E$-field is defined by the equation \cite{MacroscopicARM2021,Int_RIS_AnaSergei}:
\begin{multline}
\mathrm{E}^s_{\theta_{rm}} = \frac {jk}{4\pi}
\frac {e^{-jk|\mathbf {r}|}}{|\mathbf {r}|}E_{0}S 
\Biggr [ (\cos\theta -\cos\theta_{i})\mathrm{sinc}(ka_{\mathrm{ef}})\\
+\sum _{m}r_{m}(\theta _{i})(\cos\theta +\cos\theta _{rm})
\mathrm{sinc}(ka_{\mathrm{efm}}) \Biggr ],
\label{eq5}
\end{multline}
where $r_{m}(\theta _{i})$ is the reflection coefficient amplitude of excited harmonics and angles $\theta _{rm}$ show the deflection angles for each propagating Floquet harmonic. $|\mathbf {r}|$ is the distance from the observation point to the structure center point. $S$ and sinc$(x)$ represent the reflector area and the sinc function, respectively. Furthermore, $a_{\mathrm{ef}}$ and  $a_{\mathrm{efm}}$ are represented as $a_{\mathrm{ef}} = (\sin\theta -\sin\theta_{i})a/2$ and $a_{\mathrm{efm}} = (\sin\theta -\sin\theta_{rm})a/2$ for each reflected propagating mode. For anomalous reflectors that are perfectly optimized to reflect only into one direction, only one of the reflection coefficients $r_{m}$ is non-zero.

Comparison of the analytical results with numerical simulations for the scalable optimized panels are shown in Fig.~\ref{fig:farfield_cont}. Here, the far-field amplitude is normalized as $r|E^s(\theta)|$ calculated for the unit-amplitude incident plane wave.  We see that the performance  of the designed panels is very close to that of perfect anomalous reflectors of the same size, whose patterns are given by the analytical formula \eqref{eq5}. The differences are mainly in the side-lobe patterns: We observe that the parasitic scattering of the designed reflectors is predominantly in the specular direction and in the direction with the opposite tilt angle, while the side-lobe level is more uniform for the ideal reflector. 

\begin{table}[thb!]
\renewcommand{\arraystretch}{1.25}
    \begin{center}
    \caption{CST simulated side-lobe levels (dB) for the continuous and discrete loads for the $96\times96$ array size}
    \label{tab:SLL_Cont_discrete}
    \footnotesize
    \begin{tabular}{|c|c|c|c|c|} \hline
    \multirow{2}{4em}{Resolution}  & \multicolumn{4}{c|}{Floquet mode ($\theta^r$)} \\ \cline{2-5}
     & $1$ $(13^{\degree})$  & $2$ $(27^{\degree})$ & $3$ $(43^{\degree})$ & $4$ $(65^{\degree})$ \\ \hline
Continuous & -13.5 & -13.5 & -13.5& -13 \\ \hline
$4$~bit & -13.5 & -13.5 & -13.5 & -12.9 \\ \hline
$3$~bit & -13.5 & -13.3 & -13.4 & -9.4 \\ \hline    
$2$~bit & -13.3 & -13 & -13.4 & -8.6 \\ \hline
$1$~bit & -3.2 & -2.5 & -0.2 & 2.4 \\ \hline
    \end{tabular}
    \end{center}
\end{table}

\subsection{Quantized Load Optimization}\label{sec:RISdesign_quan}

Considering practical implementations, there are typically limitations on reaching either a wide range of load values or continuous tunability, or both. Therefore, we consider limiting the number of possible load values by quantization to mimic the digital controllability of the load reactances. The continuous phase values in Table~\ref{tab:optquantizedload} are discretized into the nearest available unit cell phase response, resulting in some phase quantization error. Specific values of the available load reactance set can be selected based on practical considerations. We utilize the standard phased-array linear phase profile and map each desired phase (shift) into the corresponding load reactance from the lookup table, built using a CST load sweep for periodically repeated single-patch unit cells at the normal incidence angle. We map the reflection phases to quantized reactive load sets for $1, 2, 3,$~and~$4$-bit resolution to have $2, 4, 8,$~and~$ 16$ distinct reactive load values, respectively.

\begin{figure*}[t!]
\begin{center}
\includegraphics[width=7in,height=1.6in]{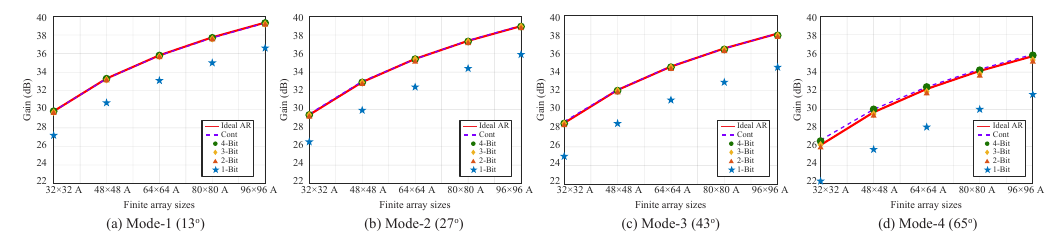}
\caption{Computed gain results CST results for anomalous reflector for continuous, $4$, $3$, $2$, \& $1$-bit quantized load sets for five different finite array sizes of $32\times32$ (blue), $48\times48$ (orange), $64\times64$ (yellow), $80\times80$ (green), and $96\times96$ (purple) for propagating harmonics from $1$ to $4$, reported in dB scale.}\label{fig:farfield_gain_Analysis}
\end{center}
\end{figure*}

The optimized outcomes with the discrete load reactance sets and the respective achievable scattering mode efficiency $\zeta$ are reported in Table~\ref{tab:optquantizedload} along with the optimized continuous load values. From the observation of simulation results, the efficiency declines for all the propagating Floquet modes when the quantization resolution decreases from $4$-bit to $1$-bit. Specifically, the performance of $1$-bit quantization demonstrates poor efficiency due to mapping continuous phase values into only two discrete phase values, raising phase quantization errors. It results in significant quantization loss, leading to reduced efficiency~\cite{YangstudyPQE}. With the moderate deflection angles for modes $1$ to $3$, the penalty for $2$-bit and $3$-bit resolution on directivity is low. With larger $\theta^r$, the quantization effect becomes slightly higher. We can conclude that the results show decent efficiencies for already $3$-bit resolution, not significantly dropping compared to continuous load optimization.  We therefore choose 3 bits for load quantization of our experimental anomalous reflector.

Further analysis on reflection efficiency, \ac{BPE}, and \ac{SLL} was conducted using the far-field radiation pattern cuts acquired for the optimized quantized loads using the array factor approximation extracted from the CST numerical solver across five distinct finite array sizes. The gain of the ideal anomalous reflector is acquired from Part~II (Eqn.~$9$). The gain results, depicted in Fig.~\ref{fig:farfield_gain_Analysis}, compare performance for continuous and quantized loads from $4$ to $1$-bit optimizations with ideal anomalous reflectors across the five array sizes for each propagating mode. Here, the gain values are calculated as directivities of the scattered field patterns (see Fig.~\ref{fig:farfield_cont}), as the panel is assumed to be lossless. The \ac{SLL} increases due to phase quantization, leading to gain loss of the optimized finite array slightly rises with the increasing propagating Floquet harmonic. However, the quantization loss increases as the phase quantization shifts from $4$ to $1$-bit across all propagating modes~\cite{YangstudyPQE}. Additionally, we observe severe deterioration for 1-bit phase quantization, and the quantization loss increases with a larger aperture array, peaking at a $96 \times 96$ array.

Table~\ref{tab:SLL_Cont_discrete} presents CST-simulated \ac{SLL} results for discrete load optimization in a $96\times96$ array. The \ac{SLL} trends remain consistent across all tested array sizes. For the used sets of quantized load reactances, the \ac{SLL} remains good for Floquet modes up to 3rd order, except for the 1-bit case. For the 4th-order mode, the side lobe level deteriorates moderately with reduced number of quantization bits. The \ac{BPE} arises only for the Floquet mode $m=4$ having a one-degree deviation from the desired direction observed in both continuous and discrete loads across $32\times32$ and $48\times48$ arrays due to small dimensions and wide-angle deflection. The \ac{BPE} increases with the increasing propagating mode index (and angle) and decreases with the expanding aperture size. The difference between the continuous load outcomes and the discretized loads arises from the restricted degrees of freedom associated with the load quantization. Significantly, the method adopted  from~\cite{vuyyuru} achieves the maximal reflection at the desired angle through an optimization process by avoiding full-wave \ac{EM} simulations. The overall objective is to develop periodic supercell models for scalable finite arrays by bypassing global optimization and replacing individual finite-sized reflector designs with algebraic optimization for a single supercell.

\section{Experimental Verification}\label{sec:RISExperimental}

A passive and fixed anomalous reflector prototype is designed, manufactured, and measured to assess the anomalous reflection performance. Based on the array factor approximation analysis derived for infinite supercell-level arrays in Section~\ref{sec:RISdesign_cont}, we choose a $48\times48$ array with a practical size for manufacturing with the available facilities and adopt the manufacturing technique outlined in~\cite{vuyyuru2023finite}. For a fair comparison between CST simulated and experimental radiation patterns, a periodic anomalous reflector antenna with updated dimensions reported in~\ref{subsec:RISFabrication} is designed and manufactured with an in-house printed circuit board (PCB) manufacturing facility as a dual-layer laser-etched PCB including galvanic through-hole plating process. An in-house EM-shielded anechoic chamber and an auditorium with dimensions $5 \times5\times2.2$~m and $14\times8\times3$~m, respectively, are used for testing the manufactured sample. The auditorium measurements are reported in Part~II.

\subsection{Practical Design and Fabrication}\label{subsec:RISFabrication}

\begin{figure*}[t]
\begin{center}
\includegraphics[width=7in,height = 4in]{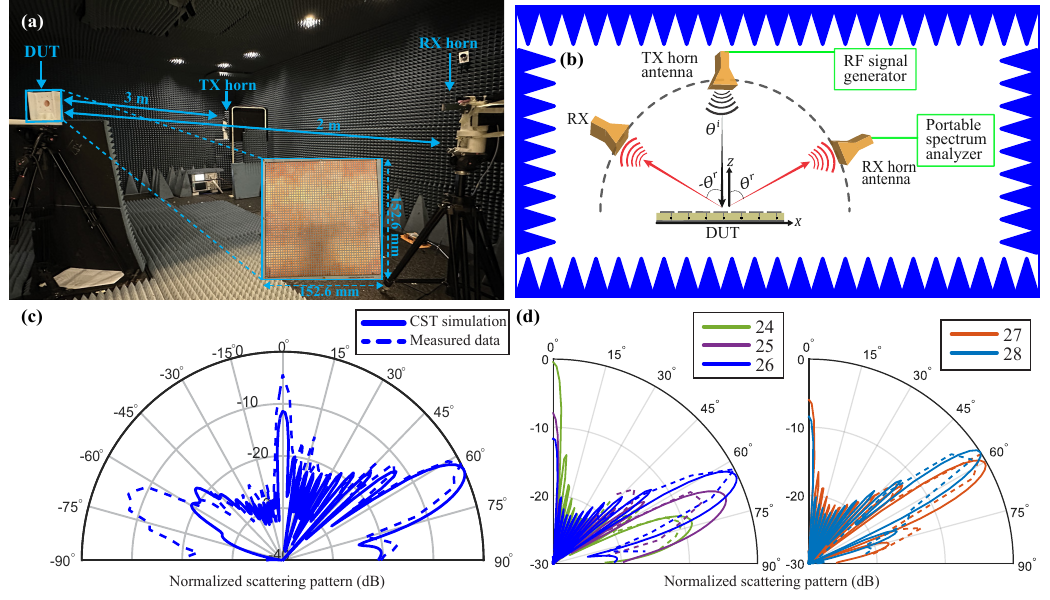}
\caption{(a) Measurement arrangement in the anechoic chamber. (b) Top view of the experimental setup. (c) Comparison of the scattering pattern results obtained through experimental measurements and numerical simulations. (d) Comparison between numerical simulations (solid line) and anechoic chamber measurement (dashed line) at several frequencies (GHz) for reflection angles from $42.5\degree$ to $82.5\degree$ with a $5\degree$ step.}
\label{fig:RIS_chamber_measurement1}
\end{center}
\end{figure*}

In Section~\ref{sec:RISdesign_cont}, we demonstrated the synthesized multi-mode anomalous reflector that can redirect an incident wave into four distinct propagation directions. Here we tweak the design a bit and present a static finite planar prototype, trying to realize the largest anomalous reflection angle (65 degrees) case of the periodic arrays. This prototype is tailored to reflect a normally incident plane wave $[(\theta^i,\phi^i)\! =\! (0,0)]$, into an anomalous reflection angle of $+65\degree$ $[(\theta^r,\phi^r)\! =\! (65\degree,0)]$ at $26$~GHz. The designed prototype array comprises $2304$ linearly (vertically) polarized individual elements arranged in a $48\times48$ square array configuration, resulting in an overall dimensions of $13.24\lambda\times13.24\lambda$.

Following~\cite{vuyyuru2023finite} for the manufacturing process, a linearly polarized copper square patch on PCB, utilizing Rogers RO4350B (LoPro) laminate with a relative permittivity $\epsilon_r\!=\!3.55$ and loss tangent $\tan(\delta)\! =\!0.0037$ with $0.52$~mm thickness is employed. We model the materials in CST as lossy and assign a surface roughness of $0.001$~mm for the copper layer. The square patch dimension is $2.385$~mm, with its feeding position located at $(x_p,y_p)\! =\! (1,0)$~mm. These dimensions are optimized to achieve resonance at $75~\Omega$ input impedance tailored for the desired frequency of $26$~GHz. This approach ensures practicality for implementing a set of optimized load reactances with shorted and open coplanar waveguide (CPW) strips. 

We constrain the phase variation to a $3$-bit (a $45\degree$ step) quantization resolution, allowing only eight distinct phase responses to be realized. This approach balances the precision and practicality of the proposed design technique in the implementation process and simplifies hardware requirements for manufacturing. Moreover, the optimization process accelerates with a reduced number of possible values for load variables.

We repeat the design procedure with the supercell comprising 16 patches using the revised unit cell geometry to synthesize the periodic array loads to reflect the normal incidence wave to the anomalous direction. The resulting far-field scattering pattern should be aligned with the extended large square finite array. The optimization of the reactive load is then executed in accordance with Section~\ref{sec:RISfloquet}. Subsequently, we scale up the initial supercell to the $48\times48$ square array.
 
Finally, in the last phase, we map the $3$-bit load reactances into a practical and equivalent set of shorted and open CPW strips, as detailed in Table~\ref{tab:CPWdimensions}. 

\begin{table} [t]
\renewcommand{\arraystretch}{1.25}
    \begin{center}
    \caption{CPW strip dimensions}
    \label{tab:CPWdimensions}
    \footnotesize
    \begin{tabular}{|c|c|c|c|c|} \hline
    No. & Phase $\degree$ & Reactances $(\Omega)$ & Strip (mm) & Term. \\   \hline
    $0$ & 0 & $106$ & $1.18$ & short \\ \hline
    $1$ & $\pi/4$ & $49$ & $0.89$ & short \\ \hline
    $2$ & $\pi/2$ & $13$ & $0.54$ & short \\ \hline
    $3$ & $3\pi/4$ & $-19$ & $1.79$ & open \\ \hline
    $4$ & $\pi$ & $-58$ & $1.44$ & open \\ \hline
    $5$ & $-3\pi/4$ & $-133$ & $1.06$ & open \\ \hline
    $6$ & $-\pi/2$ & $-583$ & $0.62$ & open \\ \hline
    $7$ & $-\pi/4$ & $294$ & $1.48$ & short \\ \hline
    \end{tabular}
    \end{center}
\end{table}

\subsection{Anechoic Chamber Measurements}\label{subsec:RISMeasurements}

The manufactured $65\degree$ reflector sample undergoes experimental testing in an anechoic chamber to carry out bistatic scattering cross-section (SCS) measurements with the device under test (DUT) securely mounted on a stationary tripod. The PCB dimensions for the anomalous reflector sample measure approximately $152.6$~mm ($x$-axis) by $152.6$~mm ($y$-axis). TX and RX antennas with a standard gain horn, offering $18.5$~dBi gain at $26$~GHz, were installed for the setup. We employ the measurement setup outlined in~\cite{vuyyuru2023finite}, where the TX horn antenna, connected to a signal generator (SG), is placed a $3$~m distance from the DUT mounted on a stationary pole. Meanwhile, the RX horn antenna, connected to a portable spectrum analyzer (SA), is at a $2$~m distance from the DUT. 

We estimate the Fraunhofer distance to be approximately $4.04$~meters, noting that the measurement is a radiating near field measurement, producing a quite reliable main lobe shape and magnitude, but less accurate side lobe data. The measurements are carried out at $26$~GHz with the receiving antenna moving at $2.5\degree$ resolution across azimuth $\in \left[ -90\degree, 90\degree \right]$. For accurate measurements in the direction of specular reflection, a 2\textdegree{} down-tilt is applied, resulting in the TX antenna at a height of 140~cm, the DUT at 129.5~cm, and the RX antenna at 122.5~cm. Due to this configuration, the incidence angle for the DUT deviates slightly from the broadside, being $[(\theta^i,\phi^i)= (2\degree,90\degree)]$. Consequently, the DUT is expected to deflect the incoming wave towards the predefined anomalous azimuth angle of $65\degree$, with a $-2\degree$ elevation. 

The normalized scattered pattern data acquired from measurements is compared with the CST simulated outcomes, as illustrated in Fig.~\ref{fig:RIS_chamber_measurement1} (c). The obtained measurement data reveals good agreement between the simulation and measurement results at the desired direction range, and shows an \ac{SLL} ratio of $-4.4$~dB. While the desired maximum measured power indicates a slight deviation by a few degrees, reflections in the specular and reverse anomalous directions are higher compared to the CST simulated data. Additionally, we studied the frequency scanning property of the anomalous reflector by measuring how the anomalous reflection angle behaves as a function of frequency. The frequency sweep data for reflection angles ranging from $42.5\degree$ to $82.5\degree$ in $5\degree$ increments is depicted in Fig.~\ref{fig:RIS_chamber_measurement1} (d), alongside numerical simulation results, demonstrating reasonable agreement between measured and simulated data. These scattering readings are normalized to the maximum power observed at 26 GHz in $62.5\degree$ direction.

From this result we can deduce that the anomalous reflection property of the designed surface is strong at least for $26$-$28$~GHz, and that increasing the frequency decreases the reflection angle, as indicated by Eq.~(\ref{eq4}). The above analysis confirms that the fixed anomalous reflector prototype performs satisfactorily at the desired frequency. The measured scattering pattern matches the main lobe with CST simulated, but the maximum peak at the desired angle is deviated due to material properties impacting the accuracy of \ac{EM} measurements.

\section{Conclusion}\label{sec:conclusion}

In this first part of a two-part submission, we design and characterize multimode periodic anomalous reflectors supporting several propagating Floquet modes through a recently developed, computationally simple array scattering synthesis method. The proposed approach efficiently deflects the incoming signals to non-specular propagating directions without requiring multiple component dimensions within a single design. The main idea involves transitioning from the comprehensive full-wave \ac{EM} simulations to a circuit-level load impedance optimization in maximizing the scattering amplitude or its propagating harmonic mode-specific reflection efficiency for the deflection toward the desired angle.

With additional design simplifications in mind, we compare optimized non-discretized load impedances with quantized load impedances down to $1$-bit resolution, demonstrating that already a $3$-bit quantization resolution decently approximates the ideal reflection performance and could be utilized in RIS \ac{EM} models allowing reduced complexity. From the quantization analysis, even $2$-bit optimized finite array designs may be suitable for large-scale implementation, even for large deflection angles, while significantly reducing system complexity and achieving efficiencies surpassing $80\%$. The experimental validation of the manufactured anomalous reflector provides empirical evidence supporting the large, wide-angle reflector design with optimized low-resolution quantized reactive loads. We address the scalability issue by avoiding design of each different finite-size reflector separately, optimizing a periodic  supercell and approximating the full array performance using the array factor. This approach seems to estimate the main scattering lobe properties of the reflector reliably. 
In future studies, the tunability of the multi-mode anomalous reflectors will be investigated, including non-perfect anomalous reflection angles between the Floquet modes' characteristic propagation angles.

In the following second part, we will transfer the developed practical technique for synthesising and modeling realistic patch arrays from \ac{EM} perspective into wireless network simulators for evaluating RIS-assisted links. We will show that the macroscopic parameters of the panels, such as gain, geometrical area, and angular settings, allow the network simulators to estimate the overall system performance and develop robust RIS-assisted wireless networks without conducting ample experiments.   

\section*{ACKNOWLEDGEMENT}

The authors express gratitude to Professor Do-Hoon Kwon from the University of Massachusetts Amherst, USA, for the valuable suggestions and discussions regarding array antenna scattering synthesis for periodic reflectors.

\ifCLASSOPTIONcaptionsoff
  \newpage
\fi

\bibliographystyle{IEEEtran}
\bibliography{IEEEabrv,Bibliography}

\end{document}

%% file: Acronyms.tex
\begin{acronym}[DSTTDSGRC]
\setlength{\itemsep}{-3pt}
\acro{EM}{electromagnetic}
\acro{SLS}{system-level simulator}
\acro{SLL}{side-lobe level}
\acro{BPE}{beam pointing error}
\end{acronym}